\documentclass[sigconf]{acmart}
\AtBeginDocument{%
  \providecommand\BibTeX{{%
    \normalfont B\kern-0.5em{\scshape i\kern-0.25em b}\kern-0.8em\TeX}}}

\usepackage{booktabs} 
\usepackage{multirow}
\usepackage{latexsym}
\usepackage{graphicx} 
\usepackage{algorithm}
\usepackage{algorithmic}
\usepackage{epsfig}
\usepackage{color}
\usepackage{amsmath}
\usepackage{graphicx,subfigure}
\usepackage{makecell}
\usepackage{textcomp}
\usepackage{url}
\usepackage{flushend}
\usepackage{bbm}

\copyrightyear{2022}
\acmYear{2022}
\setcopyright{acmcopyright}
\acmConference[SIGIR '22] {Proceedings of the 45th International ACM SIGIR Conference on Research and Development in Information Retrieval}{July 11--15, 2022}{Madrid, Spain.}
\acmBooktitle{Proceedings of the 45th International ACM SIGIR Conference on Research and Development in Information Retrieval (SIGIR '22), July 11--15, 2022, Madrid, Spain}
\acmPrice{15.00}
\acmISBN{978-1-4503-8732-3/22/07}
\acmDOI{10.1145/XXXXXX.XXXXXX}

\begin{document}
\title{News Recommendation with Candidate-aware User Modeling}

\author{Tao Qi}
\affiliation{%
  \institution{Department of Electronic Engineering,\\ Tsinghua University}
  \country{}
}
\email{taoqi.qt@gmail.com}

\author{Fangzhao Wu}
\authornote{The corresponding author.}
\affiliation{%
  \institution{Microsoft Research Asia}
  \country{}
}
\email{wufangzhao@gmail.com}

\author{Chuhan Wu}
\affiliation{%
  \institution{Department of Electronic Engineering,\\ Tsinghua University}
  \country{}
}
\email{wuchuhan15@gmail.com}

\author{Yongfeng Huang}
\affiliation{%
  \institution{Department of Electronic Engineering,\\ Tsinghua University}
  \country{}
}
\email{yfhuang@tsinghua.edu.cn}

\begin{abstract}

News recommendation aims to match news with personalized user interest.
Existing methods for news recommendation usually model user interest from historical clicked news without the consideration of candidate news.
However, each user usually has multiple interests, and it is difficult for these methods to accurately match a candidate news with a specific user interest.
In this paper, we present a candidate-aware user modeling method for personalized news recommendation, which can incorporate candidate news into user modeling for better matching between candidate news and user interest.
We propose a candidate-aware self-attention network that uses candidate news as clue to model candidate-aware global user interest.
In addition, we propose a candidate-aware CNN network to incorporate candidate news into local behavior context modeling and learn candidate-aware short-term user interest.
Besides, we use a candidate-aware attention network to aggregate previously clicked news weighted by their relevance with candidate news to build candidate-aware user representation.
Experiments on real-world datasets show the effectiveness of our method in improving news recommendation performance.

\end{abstract}



\maketitle

\section{Introduction}

Personalized news recommendation is a critical technique for online news platforms to improve user experience~\cite{yi2021efficient,qi2021uni,qi2021kim,wu2020sentirec,zheng2018drn,wuuser,khattar2018weave}.
Accurate modeling of user interest on candidate news is important for personalized news recommendation~\cite{ge2020graph,hu2020graph2,wu2020ptum,lee2020news,wu2020fairness,qi2020privacy}.
Many existing methods first model user interests and candidate news content separately and then use their representations for interest matching~\cite{wu2019neurald}.
For instance, \citet{an2019neural} used a GRU network and ID embeddings to learn user interest representations from clicked news.
\citet{wu2019ijcai} applied an attention network to learn user interest representations by aggregating user's clicked news.
Both of them modeled the relevance between user interests and candidate news based on the dot product of their representations.
In these methods, user interests are modeled in a candidate-agnostic way.
However, each user usually has multiple interests~\cite{qi2021hierec}, and it may be difficult to accurately match candidate news with a specific user interest if candidate news is not considered in user modeling~\cite{wang2018dkn}.

\begin{figure}
    \centering
    \resizebox{0.45\textwidth}{!}{
    \includegraphics{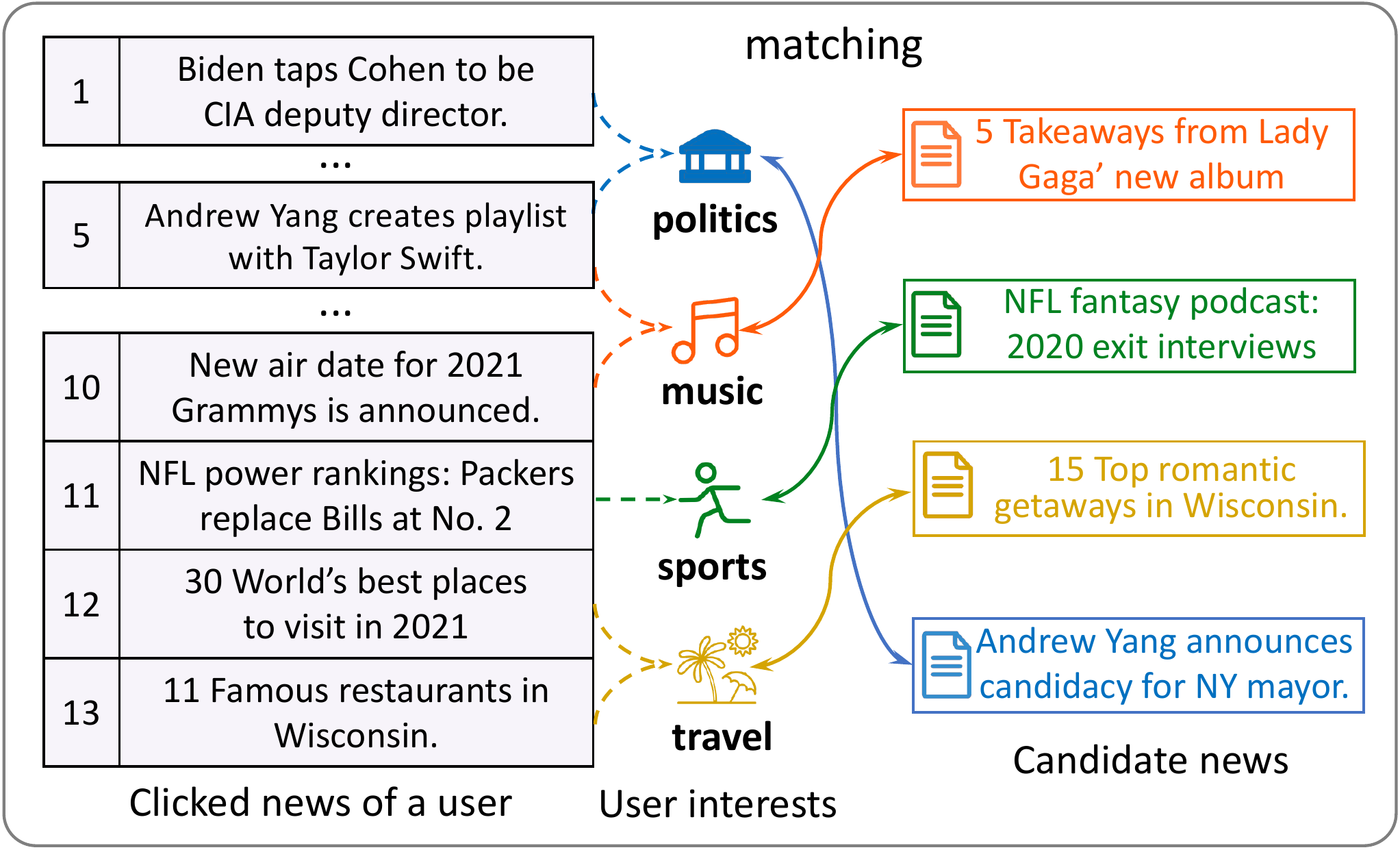}
    }
    \caption{The matching between candidate news and user interest inferred from historical clicked news.}
    \label{fig.intro}
\end{figure}

Our work is motivated by the following observations.
First, users usually have multiple interests.
For instance, as shown in Fig.~\ref{fig.intro}, we can infer that the example user is interested in many different fields, such as politics, music, sports, and travel, from her clicked news.
However, a candidate news usually only matches a small part of user interests.
For instance, the 4th candidate news only matches user interests in politics, and it has low relevance to other interests of this user like music and sports.
Thus, it may be difficult to accurately match the candidate news if candidate news information is not considered in user modeling.
Second, local contexts of users' news click behaviors are useful for inferring short-term user interests.
For example, as shown in Fig.~\ref{fig.intro}, we can infer the user's recent interests on travel in Wisconsin from the relatedness between the 12th and 13th news clicks.
Third, long-range relatedness between users' historical clicks also provides rich information to model long-term user interests.
For example, we can infer the long-term user interests in music from the long-range relatedness between the 5th and 10th clicks.
Thus, understanding both short- and long-term user interests is important for accurate news recommendation~\cite{an2019neural}.

In this paper, we propose a candidate-aware user modeling framework for personalized news recommendation (\textit{CAUM}), which can incorporate candidate news information into user modeling for accurate interest matching.
We propose a candidate-aware self-attention network to learn candidate-aware global user interest representations.
It uses candidate news representation to guide the modeling of global relatedness between historical clicked news.
In addition, we propose a candidate-aware CNN network to learn candidate-aware short-term user interest representations.
It incorporates candidate news information into the modeling of local contexts of click behaviors.
Besides, we adopt a candidate-aware attention network to weight clicked news based on their relevance with candidate news to learn candidate-aware user interest representation for better matching with candidate news.
Experimental results on two real-world datasets verify that \textit{CAUM} can improve the performance of user modeling for news recommendation.

\section{Methodology}



\subsection{Candidate-aware User Modeling}

\begin{figure}
    \centering
    \resizebox{0.47\textwidth}{!}{
    \includegraphics{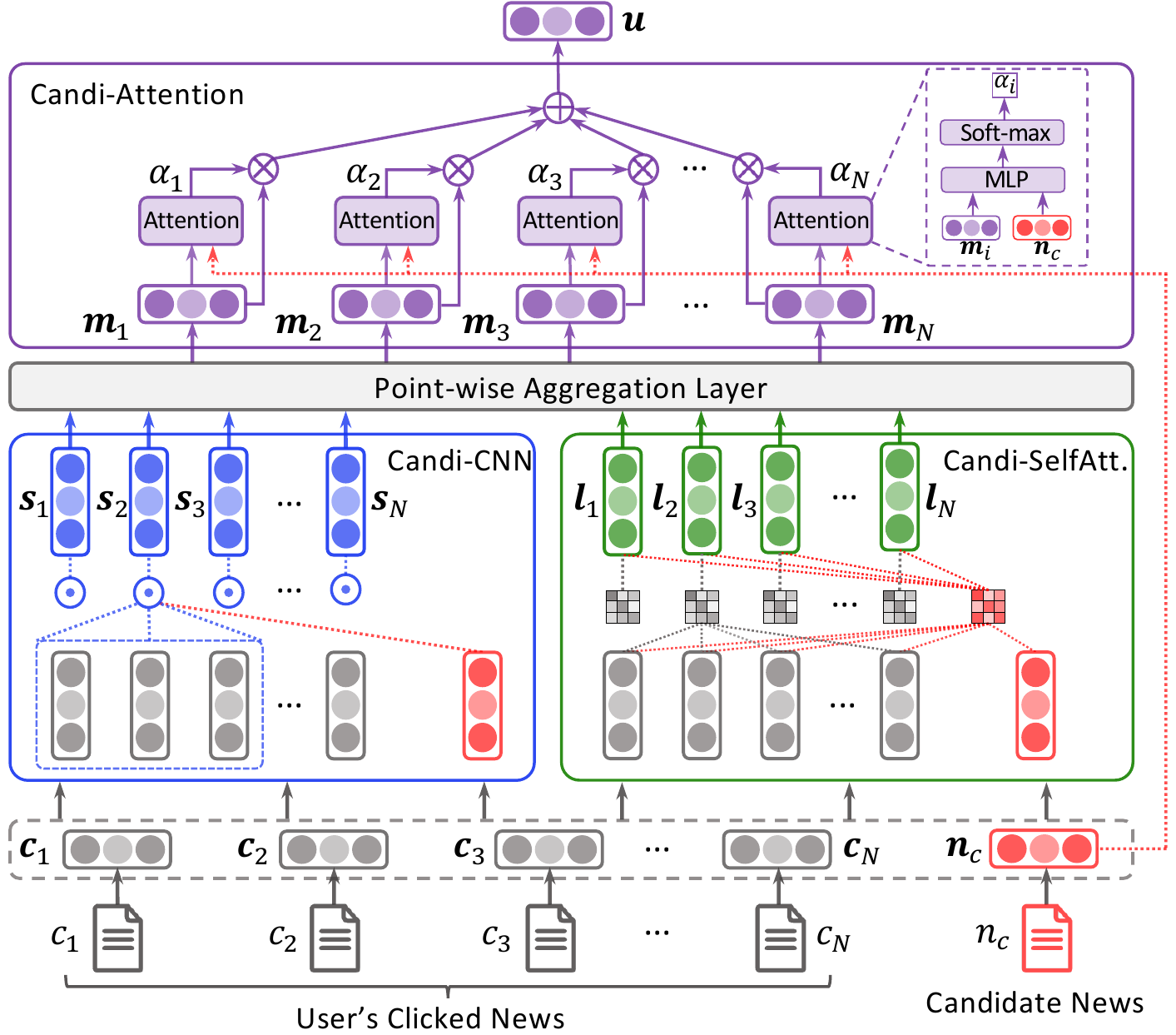}
    }
    \caption{Framework of our \textit{CAUM} method.}
    \label{fig.user_model}
\end{figure}

Next, we introduce the candidate-aware user interest modeling framework, which can exploit candidate news to guide user interests modeling.
It takes representations $\{\textbf{c}_i\}_{i=1}^N$ of user's recent $N$ clicks $\{c_i\}_{i=1}^N$ and representation $\textbf{n}_c$ of candidate news $n_c$ as inputs. (Refer to Section~\ref{sec.ne}.).
Fig.~\ref{fig.user_model} shows it contains three modules.


\textit{\textbf{Candi-SelfAtt}}:
Long-range contexts of news clicks are usually informative for inferring global user interests.
Besides, different long-range behavior contexts usually have different importance to capture different global user interests.
For example, Fig.~\ref{fig.intro} shows the relatedness between the 1st and 5th click can help infer user interests in politics while the relatedness between the 5th click and 10th click can help infer user interests in music.
Thus, modeling long-range behavior contexts with candidate news information may better model global user interests to match candidate news.
Motivated by these observations, we propose a candidate-aware self-attention network (\textit{Candi-SelfAtt}), which can use candidate news information to guide global behavior contexts modeling.
The core of \textit{Candi-SelfAtt} is to adjust attention weights of behavior contexts via candidate news to select important ones.
We first apply multiple self-attention heads~\cite{vaswani2017attention} to model click relatedness:
\begin{equation}
    \hat{r}^k_{i,j} = \textbf{q}^T_i \textbf{W}^r_k \textbf{c}_j,\quad \textbf{q}^T_i = \textbf{Q}_u\textbf{c}_i,
\end{equation}
where $\hat{r}^k_{i,j}$ denotes the attention score generated by the $k$-th attention head to model relatedness between the $i$-th and $j$-th click, $\textbf{Q}_u$ is the projection matrix, and $\textbf{W}^r_k$ is parameters of the $k$-th attention head.
Note that $\{\hat{r}^k_{i,j}\}_{j=1}^N$ model the relatedness between the $i$-th clicks and other user's clicks.
We further adaptively select informative long-range relatedness for modeling user interest in candidate $n_c$ based on their relevance with candidate news:
\begin{equation}
   r^k_{i,j} = \hat{r}^k_{i,j} + \textbf{q}^T_c \textbf{W}^r_k \textbf{c}_j, \quad \textbf{q}^T_c = \textbf{Q}_c \textbf{n}_c,
\end{equation}
where $r^k_{i,j}$ is the candidate-aware attention score, and $\textbf{Q}_c$ is a projection matrix.
Then we learn the representation $\textbf{l}^k_i$ generated by the $k$-th head for the $i$-th click based on attention weights $\{\gamma^k_j\}_{j=1}^N$:
\begin{equation}
    \textbf{l}^k_i = \textbf{W}^k_o \sum_{j=1}^N \gamma^k_j \textbf{c}_j, \ \  \gamma^k_j = \frac{\exp(r^k_{i,j})}{\sum_{p=1}^N \exp(r^k_{i,p})},
\end{equation}
where $\textbf{W}^k_o$ is the projection matrix of the $k$-th attention head.
Finally, we learn the global contextual representation $\textbf{l}_i$ for $i$-th click by contacting $\{\textbf{l}^k_i\}_{k=1}^K$, where $K$ is the number of attention heads.

\begin{table*}[]
\caption{Performance comparisons. The improvement of \textit{CAUM} over baseline methods is significant at level $p\leq0.001$.}

\centering
\resizebox{0.93\textwidth}{!}{
\begin{tabular}{ccccc|cccc}
\hline
         & \multicolumn{4}{c|}{\textit{MIND}}                                          & \multicolumn{4}{c}{\textit{NewsApp}}                                         \\ \hline
         & AUC            & MRR            & nDCG@5          & nDCG@10        & AUC            & MRR            & nDCG@5         & nDCG@10        \\ \hline
\textit{GRU}~\cite{okura2017embedding}    &65.69$\pm$0.15 &31.47$\pm$0.06 &33.96$\pm$0.07 &39.70$\pm$0.07 & 63.23$\pm$0.37 & 27.83$\pm$0.26 & 31.84$\pm$0.31 & 37.41$\pm$0.34 \\
\textit{NAML}~\cite{wu2019ijcai}    &66.49$\pm$0.19 &32.38$\pm$0.13 &35.17$\pm$0.15 &40.84$\pm$0.14 & 64.52$\pm$0.35 & 29.02$\pm$0.20 & 33.35$\pm$0.30 & 38.90$\pm$0.33 \\
\textit{NPA}~\cite{wu2019npa}      &66.56$\pm$0.18 &32.42$\pm$0.10 &35.20$\pm$0.11 &40.87$\pm$0.13 & 64.39$\pm$0.14 & 28.93$\pm$0.10 & 33.31$\pm$0.11 & 38.83$\pm$0.11 \\
\textit{NRMS}~\cite{wu2019neuralc}     &68.04$\pm$0.20 &33.31$\pm$0.07 &36.23$\pm$0.15 &41.92$\pm$0.12 & 65.36$\pm$0.28 & 29.47$\pm$0.21 & 33.96$\pm$0.27 & 39.49$\pm$0.19 \\
\textit{LSTUR}~\cite{an2019neural}    &68.36$\pm$0.22 &33.30$\pm$0.11 &36.30$\pm$0.16 &42.00$\pm$0.14 & 65.18$\pm$0.23 & 29.28$\pm$0.21 & 33.71$\pm$0.23 & 39.28$\pm$0.22 \\
\textit{KRED}~\cite{liu2020kred}      &67.73$\pm$0.13 &32.87$\pm$0.11 &35.81$\pm$0.13 &41.43$\pm$0.15 & 65.45$\pm$0.14 & 29.56$\pm$0.09 & 34.11$\pm$0.11 & 39.65$\pm$0.12 \\ \hline
\textit{DKN}~\cite{wang2018dkn}      &66.32$\pm$0.18 &32.13$\pm$0.14 &34.86$\pm$0.13 &40.47$\pm$0.18 & 62.86$\pm$0.37 & 28.00$\pm$0.23 & 32.12$\pm$0.29 & 37.68$\pm$0.28 \\
\textit{HiFi-Ark}~\cite{liu2019hi} &67.93$\pm$0.25 &32.87$\pm$0.07 &35.77$\pm$0.08 &41.47$\pm$0.10 & 64.91$\pm$0.15 & 29.10$\pm$0.12 & 33.52$\pm$0.18 & 38.98$\pm$0.14 \\
\textit{FIM}~\cite{wang2020fine}      &67.84$\pm$0.12 &33.26$\pm$0.06 &36.18$\pm$0.10 &41.86$\pm$0.11 & 65.39$\pm$0.10 & 29.63$\pm$0.11 & 34.14$\pm$0.12 & 39.60$\pm$0.10 \\
\textit{GNewsRec}~\cite{hu2020graph}   &68.36$\pm$0.22 &33.41$\pm$0.10 &36.36$\pm$0.13 &42.01$\pm$0.14 & 65.31$\pm$0.22 & 29.40$\pm$0.14 & 33.92$\pm$0.16 & 39.48$\pm$0.16 \\ 
\hline
\textit{CAUM}      &\textbf{70.04}$\pm$0.08 &\textbf{34.71}$\pm$0.08 & \textbf{37.89}$\pm$0.07 &\textbf{43.57}$\pm$0.07 & \textbf{66.44}$\pm$0.07 & \textbf{30.07}$\pm$0.10 & \textbf{34.69}$\pm$0.12 & \textbf{40.23}$\pm$0.10 \\ \hline
\end{tabular}
}
\label{table.pe}

\end{table*}

\textit{\textbf{Candi-CNN}}:
Besides global user interests, short-term user interests are also important for matching candidate news~\cite{hu2020graph,an2019neural}.
Short-term user interests can usually be effectively modeled from local contexts between adjacent user behaviors~\cite{an2019neural}.
In addition, incorporating candidate news information into local behavior contexts modeling also has the potential to better model short-term interest in candidate news.
Thus, we propose a candidate-aware CNN network, which can capture local contexts between adjacent clicks with candidate news information.
We apply multiple filters to capture the potential patterns between local contexts of adjacent clicks and candidate news: $\textbf{s}_i = \textbf{W}_c [\textbf{c}_{i-h};...;\textbf{c}_i;...;\textbf{c}_{i+h};\textbf{n}_c],$
where $\textbf{s}_i$ represents local contextual representation of the $i$-th click, $2h+1$ is the window size of the CNN network, and $\textbf{W}_c$ represents parameters of filters in the \textit{Candi-CNN} network.
Similarly, we can learn local contextual representations $[\textbf{s}_1,\textbf{s}_2,...,\textbf{s}_N]$ of all clicked news.
These local contextual representations of clicked news encode candidate-aware short-term user interests.
Then, we learn unified contextual representation $\textbf{m}_i$ for the $i$-th click based on the aggregation of $\textbf{l}_i$ and $\textbf{s}_i$: $\textbf{m}_i = \textbf{P}_m [\textbf{s}_i;\textbf{l}_i] $, where $\textbf{P}_m$ is the projection matrix.

\textit{\textbf{Candi-Att}}:
Since the importance of clicked news for modeling user interest in candidate news may be different, we apply a candidate-aware attention network to model the importance of clicked news from their relevance with candidate news $n_c$ and further build the candidate-aware user interest representation $\textbf{u}$:
\begin{equation} 
\textbf{u} = \sum_{i=1}^N \alpha_i \textbf{m}_i, \ \alpha_i = \frac{\exp(\Phi(\textbf{m}_i,\textbf{n}_c))}{\sum_{j=1}^N \exp( \Phi(\textbf{m}_j,\textbf{n}_c) )},
\end{equation}
where $\alpha_i$ is the weight of the $i$-th click and $\Phi$ is an MLP network.
In this way, user interests relevant to the candidate news can be effectively encoded into $\textbf{u}$ to improve the accuracy of interest matching.

\subsection{News Modeling}
\label{sec.ne}

In \textit{CAUM} we model news based on previous methods.
Motivated by previous works\cite{qi2021pprec,liu2020kred}, we apply self-attention networks to learn title representation $\textbf{n}^t$ and entity representation $\textbf{n}^e$ for a news $n$ from its title and entities, individually.
Besides, following \citet{wu2019ijcai}, we derive representation $\textbf{n}^v$ of news topic via a topic embedding layer.
Finally we formulate news representation $\textbf{n}$ as the aggregation of these representations: $\textbf{n} = \textbf{n}^t+\textbf{n}^e+\textbf{n}^v$.

\subsection{Interest Matching and Model Learning}

Based on the news modeling and candidate-aware user interest modeling method, we can learn representation $\textbf{n}_c$ of candidate news $n_c$, and the corresponding user representation $\textbf{u}$.
We further match them to measure user interest in candidate news and calculate the matching score: $\hat{y} = \textbf{n}^T_c \cdot \textbf{u}$.
Motivated by previous works~\cite{wu2021empowering,wu2021uag,wutanr}, we adopted BPR loss~\cite{rendle2009bpr} for model learning:$\mathcal{L} = -\frac{1}{H}\sum_{i=1}^H \log\phi(\hat{y}^p_i-\hat{y}^n_i)$, where $H$ is the training dataset size, $\phi$ is sigmoid function, $\hat{y}^p_i$ and $\hat{y}^n_i$ is the matching score of the $i$-th positive and negative sample, and negative samples are randomly sampled for each positive sample from the same impression.

\section{Experiment}

\subsection{Dataset and Experimental Settings}

We conduct extensive experiments on two real-world datasets.
The first one is a public news recommendation dataset (\textit{MIND})~\cite{wu2020mind}.
The second one is \textit{NewsApp}, consisting of user logs collected from the news feeds app of Microsoft from January 23 to April 01, 2020 (13 weeks).
It contains 110,000 and randomly selected from the first ten weeks for training, and 100,000 impressions randomly selected from the last three weeks to construct the test set.



The data processing of \textit{CAUM} follows \citet{wu2019neuralc}.
In \textit{CAUM}, dimensions of both news and user interest representations are set to 400.
\textit{Candi-SelfAtt} contains 20 attention heads, and output vectors of each head are 20-dimensional.
\textit{Candi-CNN} contains 400 filters and window size is set to 3.
We train \textit{CAUM} 3 epochs via Adam~\cite{kingma2014adam} with $5\times10^{-5}$ learning rate.
All hyper-parameters of \textit{CAUM} and other baseline methods are selected based on the validation dataset.
Codes are in \url{https://github.com/taoqi98/CAUM}.
Following \citet{wu2020mind}, we adopted AUC, MRR, nDCG@5, and nDCG@10 for evaluation.

\subsection{Performance Comparison}

We compare \textit{CAUM} with several SOTA baseline methods:
(1) \textit{GRU}~\cite{okura2017embedding}: model user interest via a GRU network.
(2) \textit{DKN}~\cite{wang2018dkn}: apply a candidate-aware attention network to learn user representation.
(3) \textit{NAML}~\cite{wu2019ijcai}: learn user representation via a attention network.
(4) \textit{NPA}~\cite{wu2019npa}: propose a personalized attention network to model user interests.
(5) \textit{HiFi-Ark}~\cite{liu2019hi}: learn user representation from multiple archives of user interests via a candidate-aware attention network.
(6) \textit{LSTUR}~\cite{an2019neural}: use GRU network and ID embeddings to model long- and short- term user interests.
(7) \textit{NRMS}~\cite{wu2019neuralc}: apply self-attention network to model user interests.
(8) \textit{KRED}~\cite{liu2020kred}: model news from title and entities via a KGAT network.
(9) \textit{GNewsRec}~\cite{hu2020graph}: adopt GRU and GNN network to model user interests.
(10) \textit{FIM}~\cite{wang2020fine}: utilize 3D-CNN to model user interests in candidate news based on word-level similarities between clicks and candidate.

We repeat each experiment 5 times and report average results and standard deviations in Table~\ref{table.pe}.
First, we find that \textit{CAUM} can significantly outperform other baseline methods which model user interests in a candidate-agnostic manner, such as \textit{NRMS} and \textit{LSTUR}.
This is because users usually have multiple interests and a candidate news usually only matches a specific user interest.
Modeling user interests in a candidate-agnostic manner makes these methods cannot effectively capture user interests that are relevant to a specific candidate news, and maybe sub-optimal for the interest matching.
Second, our \textit{CAUM} method also outperforms baseline methods with candidate-aware attention network, such as \textit{DKN}, \textit{HiFi-Ark} \textit{GNewsRec}.
This is because different contexts of user's news clicks usually contain various clues to infer different user interests.
Incorporating candidate news information into the behavior contexts modeling can help capture more relevant user interests for matching the candidate news.
However, in these methods, user behavior contexts are ignored (e.g., \textit{DKN}) or modeled in a candidate-agnostic way (e.g., \textit{HiFi-Ark}), which are sub-optimal for the interest matching.
Different from these methods, we propose \textit{Candi-SelfAtt} and \textit{Candi-CNN} to exploit candidate news to guide the user modeling from both long-range and local behavior contexts.

\begin{figure}
    \centering
    \resizebox{0.46\textwidth}{!}{
    \includegraphics{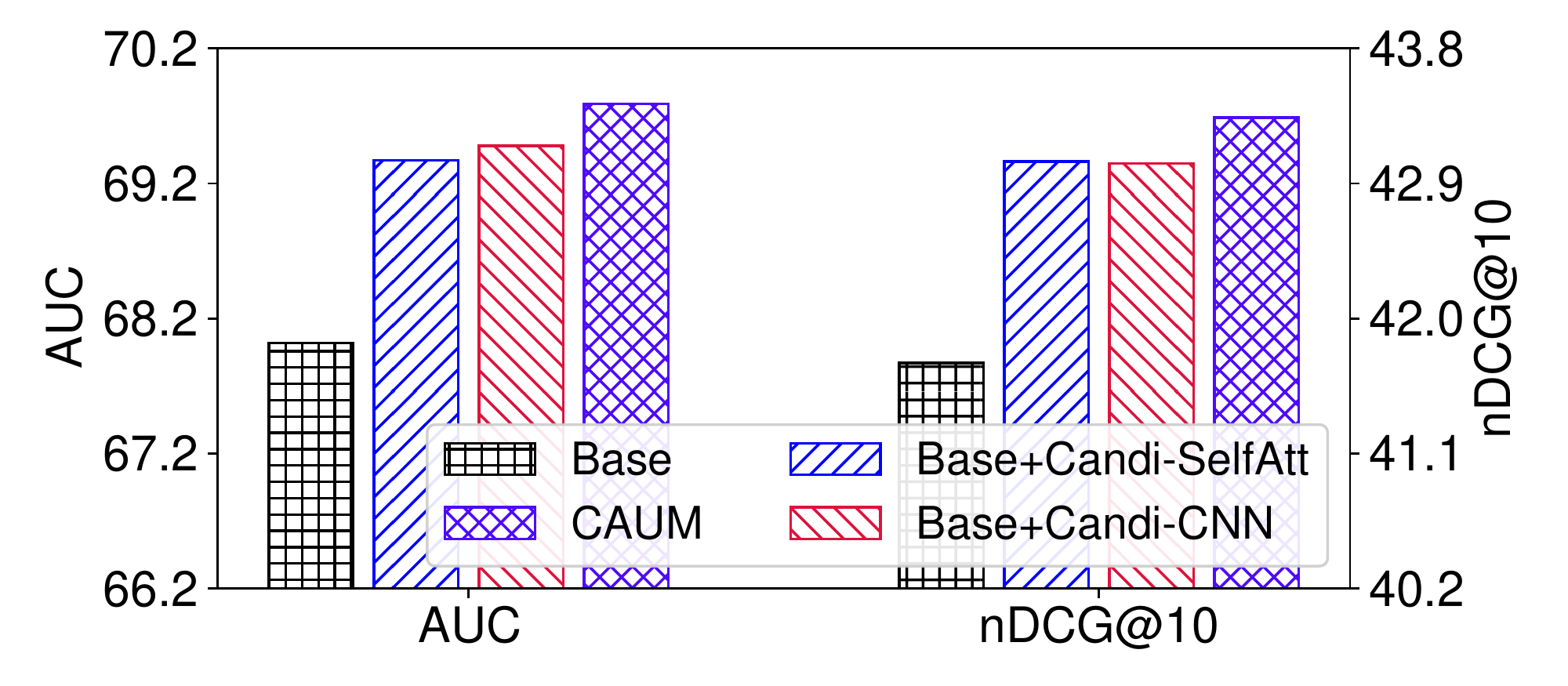}
    }
    \caption{Ablation study of \textit{CAUM}.}
    \label{fig.ablation}
\end{figure}

\begin{figure}
    \centering
    \resizebox{0.49\textwidth}{!}{
    \includegraphics{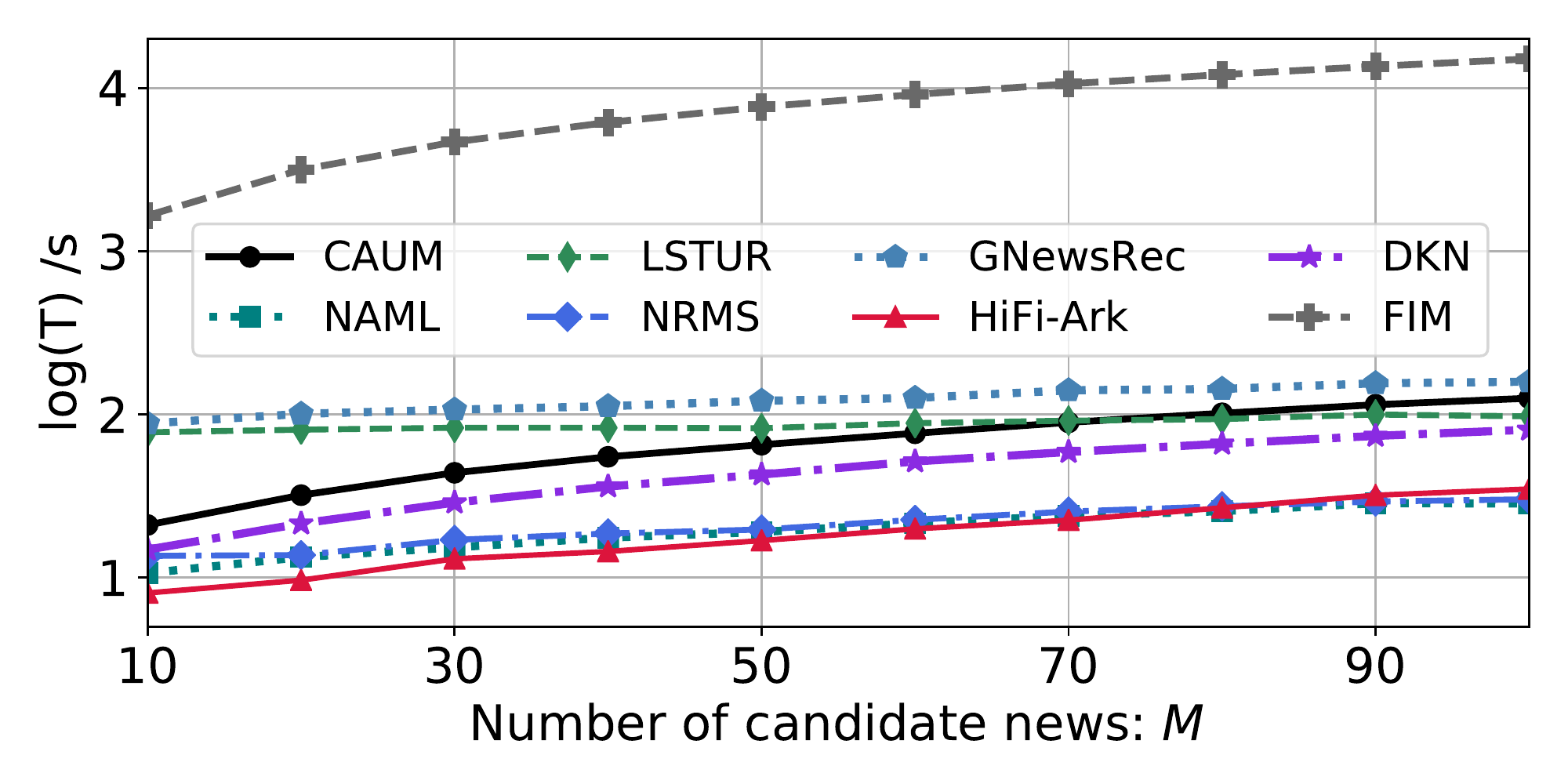}
    }
    \caption{Normalized inference time of different methods.}
    \label{fig.time}
\end{figure}

\subsection{Ablation Study}

We conduct an ablation study to verify the effectiveness of \textit{Candi-SelfAtt} and \textit{Candi-CNN} by adding them to the base model of \textit{CAUM} (named \textit{Base}).
\textit{Base} is a variation of \textit{CAUM} that replaces \textit{Candi-SelfAtt}, \textit{Candi-CNN} and \textit{Candi-Att} network with self-attention, CNN, and attention network individually.
Results are shown in Fig.~\ref{fig.ablation} and we have several findings.
First, adding one of \textit{Candi-SelfAtt} and \textit{Candi-CNN}  significantly improves the performance of \textit{Base}.
This result verifies that \textit{Candi-SelfAtt} and \textit{Candi-CNN} can effectively exploit behavior contexts to capture global and short-term user interests that are informative for matching the candidate news, respectively.
Second, \textit{CAUM} outperforms both \textit{Base+CandiCNN} and \textit{Base+Candi-SelfAtt}.
This is because candidate-aware user interest modeling is important for both global and short-term user interest.

\subsection{Analysis on Model Efficiency}

We will present some efficiency analysis and comparisons on \textit{CAUM} and other user modeling methods.
First, in Table~\ref{table.time}, we show time complexities of \textit{CAUM} and candidate-agnostic methods for calculating matching scores of $M$ candidate news for a user.\footnote{These methods can directly exploit news representations calculated in advance.}
A notable result is that although \textit{CAUM} needs to calculate different user representations for different candidate news, the time complexity of \textit{CAUM} is not $M$ times that of other methods.
This is because, in \textit{CAUM}, many operations only need to be performed once for different candidate news such as calculating self-attention scores $\hat{r}^k_{i,j}$ between different clicked news.
Thus, by avoiding executing duplicated calculations, the efficiency of \textit{CAUM} can be significantly improved.
Besides, in general, the number of candidate news $M$ is usually in a small scale (e.g., 100) in real-world recommender systems and it is comparable with the number of users' clicked news $N$ used for interest modeling (e.g., 50).
Thus, in practical settings, \textit{CAUM} can achieve comparable time complexity with \textit{NRMS}.
In addition, although \textit{GRU} and \textit{LSTUR} have smaller time complexity than \textit{NRMS} and \textit{CAUM}, it is difficult to speed up these RNN based methods via parallel computations and they usually cost more time in real applications.
Second, as shown in Fig.~\ref{fig.time}, we compare running time $T$ of different methods for calculating matching scores of $M$ candidate news for 100,000 users.
Different methods are executed in the same experimental environment (a Nvidia 1080 Ti GPU).
We find that \textit{CAUM} can achieve comparable speeds with many candidate-agnostic methods (e.g., \textit{NAML} and \textit{NRMS}) and outperform some candidate-agnostic methods (e.g., \textit{LSTUR}).
These results further verify that the efficiency of \textit{CAUM} is satisfied like candidate-agnostic methods.

\begin{table}[]
\caption{Method time complexity (multiplication operation) of calculating matching scores of $M$ candidate news. News and user representation are $d$-dimensional.}

\resizebox{0.48\textwidth}{!}{
\begin{tabular}{|c|c|c|c|}
\hline
NAML  &  $\mathcal{O}(Md + Nd^2)$    & GRU      & $\mathcal{O}(Md + Nd^2)$     \\ \hline
LSTUR & $\mathcal{O}(Md + Nd^2)$      & NRMS     &  $\mathcal{O}(3Nd^2 + N^2d+Md)$   \\ \hline
CAUM  & \multicolumn{3}{c|}{ $\mathcal{O}((3N+M)d^2 + (N^2+MN)d)$ } \\ \hline
\end{tabular}
}
\label{table.time}
\end{table}

\section{Conclusion}

In this paper, we propose a candidate-aware user modeling framework for personalized news recommendation (\textit{CAUM}), which can incorporate candidate news information into user modeling for more accurate interest matching.
We propose a candidate-aware self-attention network to exploit candidate news information as clue to model global user interests in candidate news.
In addition, we also propose a candidate-aware CNN network to incorporate candidate news information into local click behavior contexts modeling to match short-term user interests with the candidate news.
Extensive experiments on two real-world datasets demonstrate that \textit{CAUM} can significantly outperform many baseline methods and improve the accuracy of user modeling.


\section*{Acknowledgments}
This work was supported by the National Natural Science Foundation of China under Grant numbers 2021ZD0113902, U1936208, and U1936216.
We are grateful to the reviewers for their great comments and suggestions on this work.

\bibliographystyle{ACM-Reference-Format}
\bibliography{mybib}


\end{document}